\documentclass[preprint,showpacs,preprintnumbers,amsmath,amssymb]{revtex4}
\usepackage{graphicx}
\usepackage{dcolumn}
\usepackage{bm}

\begin{document}

\title{On the Decay Rate of the False Vacuum \\}

\author{ Marco Zoli }
\affiliation{Istituto Nazionale Fisica della Materia -
Dipartimento di Fisica
\\ Universit\'a di Camerino, 62032, Italy. - marco.zoli@unicam.it}

\date{\today}

\begin{abstract}
The finite size theory of metastability in a quartic potential is
developed by the semiclassical path integral method. In the
quantum regime, the relation between temperature and classical
particle energy is found in terms of the first complete elliptic
integral. At the sphaleron energy, the criterion which defines the
extension of the quantum regime is recovered. Within the latter,
the temperature effects on the fluctuation spectrum are evaluated
by the functional determinants method and computed. The eigenvalue
which causes metastability is determined as a function of
size/temperature by solving a Lam\`{e} equation. The ground state
lifetime shows remarkable deviations with respect to the result of
the infinite size theory.
\end{abstract}


\maketitle

\section*{I. Introduction  }

The theory of tunneling in a metastable potential has been a
widely investigated topic after the seminal works by Langer and
Coleman which provided the mathematical basis of the decay
processes in statistical physics \cite{langer} and quantum field
theory \cite{cole}. In the latter the decay rate of false ground
states is suitably evaluated by the Euclidean path integral method
in the semiclassical approximation \cite{laughlin,miller}: the
classical particle paths are selected as the solutions of the
Euler-Lagrange equations fulfilling some boundary conditions and,
around such stationary points, the imaginary time action is
expanded up to second order in the quantum fluctuations whose
contribution is evaluated by means of the theory of the functional
determinants. It is in the spectrum of the fluctuations that lie
the origins of metastability \cite{schulman}. To be specific, take
a particle of mass $M$ moving in a nonlinear potential with
frequency $\omega$ and negative quartic parameter $\delta$:

\begin{eqnarray}
V(x)=\, {{M\omega^2} \over 2}x^2 - {{\delta} \over 4}x^4
\label{eq:0}
\end{eqnarray}

\begin{figure}
\includegraphics[height=7cm,angle=-90]{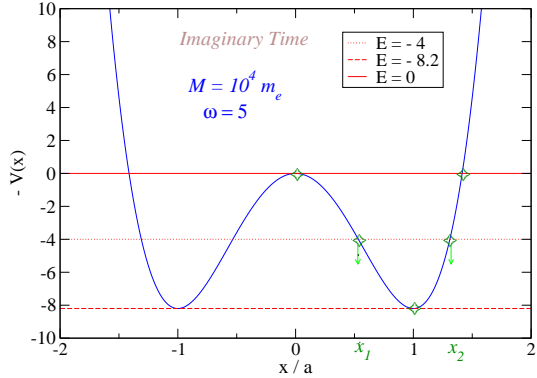}
\caption{\label{fig:1}(Color online) Reversed potential of
Eq.~(\ref{eq:0}) (units meV). $x_1$,$x_2$ are the points in which
the classical path velocity vanishes. Their location varies with
$E$ (in meV). $\omega$ is in meV.}
\end{figure}

Saying $\pm a$ are the positions of the potential maxima,
$\delta$ is given by $\delta=\, {M\omega^2/ a^2}$ in units $eV
\AA^{-4}$. Let's set $a=\,1\AA$ throughout the paper. The point
$x=\,0$ is a classical ground state but quantum mechanically the
particle can penetrate the hills and explore the abysses at
$|x|\geq \sqrt{2} a$. To see how it happens, let's focus on the
positive $x-$axis and consider the classical problem after
performing a Wick rotation which maps the time from the real to
the imaginary axis, $t \rightarrow -i\tau$, and turns the
potential upside down as shown in Fig.~\ref{fig:1}.

Now the classical equation of motion admits a non trivial solution
which, in the {\it infinite size} formalism, is known as the {\it
bounce}. In the path integral language, the {\it bounce} is given
by that path that starts at $(x_1=\,0, \, \tau=\,-\infty)$,
reaches the escape point at $(x_2=\,\sqrt{2}a, \, \tau=\,\tau_0)$
and bounces back at $(x_1=\,0, \, \tau=\,+\infty)$. The canonical
bounce solution is therefore $\tau-$ reversal invariant and
consistent with the zero energy $E$ motion (see Fig.~\ref{fig:1})
between the turning points $x_1,\,x_2$. In the latter the path
velocity vanishes. As the bounce path has a maximum versus $\tau$
the bounce path velocity has a node, hence the zero mode of the
Schr\"{o}dinger-like stability equation which governs the quantum
fluctuations cannot be the ground state \cite{affl}. In other
words, the bounce is a saddle and not a minimum for the Euclidean
action. Thus, there must be a fluctuation whose eigenvalue is
lower in energy than the zero mode eigenvalue. It is this negative
eigenvalue which requires an analytic continuation in the Gaussian
integral and ultimately leads to an imaginary square root
fluctuation determinant. When multi bounce solutions are taken
into account (due to multiple excursions between $x_1,\,x_2$) one
finally gets an explicit formula for the semiclassical tunneling
rate as given by the imaginary part of the ground state energy.
The fact is that such tunneling rate does not depend on the
temperature as the whole theory is based on the {\it infinite
size}, or $E=\,0$, formalism. The questions I want to deal with in
this paper are the following: how is the quantum fluctuations
spectrum affected by temperature effects in the quantum regime?
And, accordingly, to which extent is the particle lifetime
shortened? To get quantitative answers a {\it finite size} theory
of metastability has to be developed. The focus is here on a non
dissipative system \cite{grabert,legg}.

In Section II, I select the family of classical paths which makes
stationary the Euclidean action thus generalizing the bounce
solution for finite $E$. In Section III, the path integral of the
metastable potential is presented together with the solution of
the stability equation which governs the quantum fluctuations. The
tunneling rate for the {\it finite size} theory is calculated in
Section IV while the conclusions are drawn in Section V.

\section*{II. Finite Size Classical Bounce  }

In the imaginary time formalism \cite{jackiw} the particle motion
is classically allowed in $-V(x)$ (Fig.~\ref{fig:1}) as the
equation of motion reads:

\begin{eqnarray}
M\ddot{x}_{cl}(\tau)=\,V'(x_{cl}(\tau)) \label{eq:1}
\end{eqnarray}

where $V'$ means derivative with respect to $x_{cl}$. Let's define

\begin{eqnarray}
& &\chi_{cl}(\tau)=\, {1 \over {\sqrt{2}}}{{x_{cl}(\tau)} \over a}
\, \nonumber
\\
& &\kappa=\,{{2E}\over {\delta a^4}} \label{eq:1a}
\end{eqnarray}

and integrate Eq.~(\ref{eq:1}). Then, one gets:

\begin{eqnarray}
& &\tau - \tau_0 =\,\pm {{1 \over \omega}}
\int_{\chi_{cl}(\tau_0)}^{\chi_{cl}(\tau)} {{d\chi} \over
{\sqrt{-\chi^4 + \chi^2 + \kappa/2 }}} \, \nonumber
\\
\label{eq:4}
\end{eqnarray}

with integration constant $E$ representing the classical energy
associated to the particle motion. In Fig.~\ref{fig:1}, the
potential barrier is sufficiently high to make the semiclassical
approximation valid. The center of motion can be set at
$\tau_0=\,0$ with no loss of generality, then $\tau \in [-L/2,
L/2]$ and $L$ is the {\it size} of the system, that is the period
for a particle excursion to the abyss and back to the starting
position. Then the {\it size} represents the finite time required
for this journey to occur. For negative energies, $- {{\delta a^4}
\over 4} < E \leq 0$, there are two turning points $x_1 \,
(\chi_1)$ and $x_2 \, (\chi_2)$ ($0 \leq x_1 \leq x_2$) which
define the bounds for the particle excursion. They are given by:

\begin{eqnarray}
\chi_1=\, \sqrt{{1 - \sqrt{1 + 2\kappa}} \over 2} \,{} ; {} \,
\chi_2=\, \sqrt{{1 + \sqrt{1 + 2\kappa}} \over 2} \label{eq:5}
\end{eqnarray}

\begin{figure}
\includegraphics[height=7cm,angle=-90]{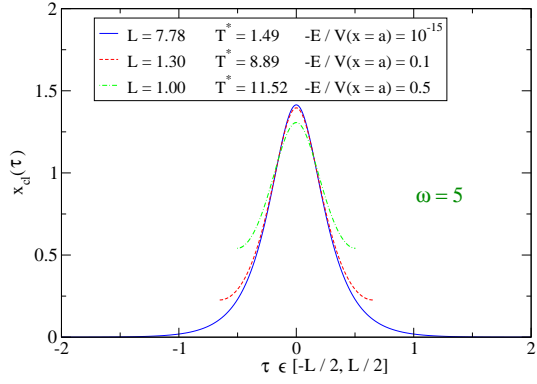}
\caption{\label{fig:2}(Color online) Shape of the bounce solution
for various classical energies $E$. $L$ is in $sec$ and $T^*$ in
$K$.}
\end{figure}

Thus the {\it length} of the generalized bounce, $\chi_2 -
\chi_1$, depends on $E$ and shrinks from 1 ($E=\,0$) to 0
($E=\,-V(x=\,a)$) \cite{n1}. This marks an essential difference
with respect to the bistable $\phi^4$ model \cite{i1} in which the
(anti)instantons interpolates between the two potential minima
thereby covering a distance that does not depend on $E$. To
integrate Eq.~(\ref{eq:4}) I use the result \cite{grad}

\begin{eqnarray}
& & \int_{\chi_{cl}(\tau)}^{\chi_2}{d\chi \over {(\chi^2 -
\chi_1^2)(\chi_2^2 - \chi^2)}}=\, {1 \over \chi_2}F(\zeta, m)\,
\nonumber
\\
& & \zeta=\,\arcsin\Biggl[\sqrt{{\chi_2^2 - \chi_{cl}(\tau)^2}
\over {\chi_2^2 - \chi_1^2}}\Biggr] \, \nonumber
\\
& & m^2=\, 1 - \chi_1^2/\chi_2^2 \label{eq:8}
\end{eqnarray}

with $F(\zeta,m)$ being the elliptic integral of the first kind
with amplitude $\zeta$ and modulus $m$. Then, after setting
$\chi_{cl}(\tau_0)=\,\chi_2$, from
Eqs.~(\ref{eq:1a}),~(\ref{eq:4}),~(\ref{eq:8}) I get the
generalized bounce for the {\it finite size} theory:

\begin{eqnarray}
& &x_{cl}(\tau)=\, a\sqrt{2}{{\chi_2}{dn(\varpi,m)}} \, \nonumber
\\
& &\varpi=\,{{\chi_2 \omega (\tau - \tau_0)} } \label{eq:9}
\end{eqnarray}

where ${dn(\varpi,m)}$ is the Jacobi delta-amplitude defined in
Eq.~(\ref{eq:55}). In the $E=\,0$ limit, $m^2=\,1$, hence one
recovers the bounce solution of the infinite size theory,
$x_{cl}(\tau)=\,a\sqrt{2} \cosh^{-1}(\varpi)$. Fig.~\ref{fig:2}
plots some paths of the family in Eq.~(\ref{eq:9}): the shape of
the paths is not essentially modified (with respect to the $E
\simeq 0$ path) up to energies of order $|E| \simeq 0.1V(x = a)$.
Above this value the bounce progressively shrinks and becomes a
point-like object at the sphaleron energy \cite{kuznet} defined by
$E_{sph}\equiv\,|E|=\,V(x = a)$.

$dn(\varpi,m)$ is even function of $\varpi$ in the period
$[-K(m),K(m)]$ determined by the first complete elliptic integral
$K(m)=\, F(\pi/2,m)$ \cite{abram}. Then, imposing the boundary
conditions on the $\tau-$range, I get from Eq.~(\ref{eq:9}) (with
$\tau_0=\,0$):

\begin{eqnarray}
\omega L=\,{{2K(m)}\over {\chi_2}} \label{eq:11}
\end{eqnarray}

Eq.~(\ref{eq:11}) expresses the fundamental relation between the
{\it size} $L$ and the particle energy $E$ which is hidden in the
modulus according to Eqs.~(\ref{eq:1a}),~(\ref{eq:5}) and the last
of Eq.~(\ref{eq:8}). Mapping $L$ onto the temperature axis,
$L=\,\hbar /(K_BT^*)$ in the spirit of the thermodynamic Bethe
Ansatz \cite{zamol}, one obtains the link between $E$ and the
temperature $T^*$ at which the particle motion takes place:

\begin{eqnarray}
K_BT^*=\,{{\hbar \omega} \over 2} {{\chi_2} \over { K(m)}}
\label{eq:110}
\end{eqnarray}

For $E=\,-V(x=\,a)$, $K(m=\,0)=\,\pi/2$ hence, Eq.~(\ref{eq:110})
leads to the Goldanskii criterion \cite{gold} for the transition
between quantum and activated regime: $K_BT_{max}^*=\,\hbar\omega
/(\sqrt{2}\pi)$.  $T_{max}^*$ is thus the maximum temperature at
which the tunneling occurs. Computation of Eq.~(\ref{eq:11}) shows
that $L(|E|)$ is a monotonically decreasing function below the
sphaleron energy. This confirms, on general grounds \cite{chudno},
that the transition at $T_{max}^*$ is expected to be a smooth
crossover as proposed long time ago \cite{affl,larkin}. The
physical origin of the smooth change lies in the fact that the
bounce is flexible and continuously adapts its shape to the
periodic and changeable (with $E$) boundary conditions. Thus the
physical picture differs very much from that encountered in
studying the bistable $\phi^4$ potential \cite{i1,i2,schaefer}
where the (anti)instantons have to fulfill antiperiodic boundary
conditions which are simply imposed by the potential structure. As
a consequence the quantum/activated crossover for the bistable
$\phi^4$ potential is in fact a sharp transition \cite{i1}.

The same conclusion regarding the character of the transition at
$T_{max}^*$ for the metastable $\phi^4$ potential may be drawn by
a direct evaluation of the classical action $A[x_{cl}]$ which, for
the {\it finite size} bounce, reads

\begin{eqnarray}
A[x_{cl}]=\, 2\sqrt{2M} \int_{x_{1}}^{x_{2}}dx \sqrt{E + V(x)} -
E\cdot L \label{eq:40}
\end{eqnarray}

Using Eqs.~(\ref{eq:0}),~(\ref{eq:9}),~(\ref{eq:11}) one can
monitor the smooth $E$ (or $T^*$) dependence of $A[x_{cl}]$ and
its derivative \cite{blatter}. For $E\rightarrow\,0$,
Eq.~(\ref{eq:40}) leads to the {\it infinite size} result

\begin{eqnarray}
A[x_{cl}]/\hbar \rightarrow {{4 M^2 \omega^3} /
({3\hbar\delta}})\label{eq:401}
\end{eqnarray}

The fact that $A[x_{cl}] \propto 1/\delta$ represents the
motivation for the semiclassical approach to the quantum
tunneling.

\section*{III. Semiclassical Path Integral }

The particle path can be written as a sum of the classical
background and the quantum fluctuation, $x(\tau)=\,x_{cl}(\tau) +
\eta(\tau)$. Then, the semiclassical space-time particle
propagator between the positions $x_i$ and $x_f$ in the imaginary
time $L$ is given, in quadratic approximation, by

\begin{eqnarray}
& &<x_f|x_i>_L=\,\exp\biggl[- {1 \over {\hbar}} A[x_{cl}] \biggr]
\cdot \int D\eta \exp\biggl[- {1 \over {\hbar}} A_f[\eta] \biggr]
\, \nonumber
\\
& &A_f[\eta]=\,\int_{-L/2}^{L/2} d\tau {M \over 2}\biggl(
\dot{\eta}^2(\tau) + {1 \over M}V''(x_{cl}(\tau))\eta^2(\tau)
\biggr) \, \nonumber
\\
& &{{V''(x_{cl}(\tau))}}=\,M{\omega^2} \bigl(1 - 6\chi_2^2
dn^2(\varpi,m) \bigr) \label{eq:25}
\end{eqnarray}

As $x_i$ and $x_f$ coincide in our model, Eq.~(\ref{eq:25}) is the
single bounce contribution $Z_1$ to the total partition function
which determines the tunneling rate. Let's evaluate it. The
measure $D\eta$ of the path integration is given through the
coefficients $\varsigma_n$ of the fluctuation expansion in a
series of ortonormal components $\eta_n(\tau)$:

\begin{eqnarray}
& &\int D\eta=\,\aleph \prod_{n=\,0}^{\infty}
\int_{-\infty}^{\infty} {{d\varsigma_n}\over
{\sqrt{2\pi\hbar/M}}}\, \nonumber \\ &
&\eta(\tau)=\,\sum_{n=\,0}^{\infty} \varsigma_n \eta_n(\tau)
\label{eq:34}
\end{eqnarray}

The normalization constant $\aleph$ accounts for the Jacobian in
the transformation to the normal mode expansion and the components
$\eta_n(\tau)$ \cite{n3} are the eigenstates (whose eigenvalues
are denoted by $\varepsilon_n$) of the Schr\"{o}dinger-like
equation:

\begin{eqnarray}
\Bigl[-{{d^2} \over {d\tau^2}} + {{V''(x_{cl}(\tau))} \over M}
\Bigr] \eta_n(\tau) =\,\varepsilon_n  \eta_n(\tau) \label{eq:26}
\end{eqnarray}

Supposed to have solved Eq.~(\ref{eq:26}) (see below), after
performing Gaussian integrations over the {\it directions}
$\varsigma_n$, one formally obtains the quantum fluctuations
factor:

\begin{eqnarray}
& &\int D\eta \exp\biggl[- {1 \over {\hbar}} A_f[\eta]
\biggr]=\,{\aleph \Bigl(Det[\hat{O}] \Bigr)^{-1/2}} \, \nonumber
\\
& & Det[\hat{O}]\equiv \, \prod_{n=\,0}^{\infty}\varepsilon_n
\,;{} \, \hat{O}\equiv \,-\partial_{\tau}^2 + {{ V''(x_{cl})}\over
M} \label{eq:35}
\end{eqnarray}

There is however a trouble in Eq.~(\ref{eq:35}) due to the zero
eigenvalue $\varepsilon_0$ which breaks the Gaussian approximation
making $Det[\hat{O}]^{-1/2}$ divergent. It can be easily seen
\cite{schulman} that the zero mode {\it is} \cite{n4} the path
velocity $\dot{x}_{cl}(\tau)$ which solves the Euler-Lagrange
problem and the homogeneous differential equation associated to
the Eq.~(\ref{eq:26}). The physical origin of the zero mode lies
in the fact that the center of motion $\tau_0$ of the periodic
bounce solutions can be placed everywhere in the range
$[-L/2,L/2]$. This mirrors the $\tau$-translational invariance of
the system. Then the zero eigenvalue can be extracted from
$Det[\hat{O}]$ and, resorting to the integration over
$\varsigma_0$ in Eq.~(\ref{eq:34}), it can be replaced in
Eq.~(\ref{eq:35}) according to the recipe: $(\varepsilon_0)^{-1/2}
\rightarrow \sqrt{{{A[x_{cl}]}\over {2\pi\hbar}}}L$
\cite{schulman,n4a}.

\section*{A. Functional Determinant }

Now I proceed invoking the theory of the functional determinants
\cite{gelfand,forman,cole1,kirsten1,burgh,lesch} to compute the
whole fluctuation contribution embodied in the regularized
determinant $Det^R[\hat{O}]$ such that $Det[\hat{O}]\equiv
\varepsilon_0 \cdot Det^R[\hat{O}]$. The form of $Det^R[\hat{O}]$
depends on the type of boundary conditions fulfilled by the
quantum fluctuations $\eta_n(\tau)$ and their derivatives. To
establish it, remember that $\dot{x}_{cl}(\tau)$ {\it is} a
quantum fluctuation whose periodicity can be promptly checked by
deriving Eq.~(\ref{eq:9}):

\begin{eqnarray}
& &\dot{x}_{cl}(\tau)=\,-a\omega \sqrt{2}m^2 \chi_2^2
sn(\varpi,m)cn(\varpi,m) \, \nonumber
\\
& &\ddot{x}_{cl}(\tau)=\,-a\omega^2 \sqrt{2}m^2 \chi_2^3
dn(\varpi,m)\bigl[1 - 2sn^2(\varpi,m)\bigr] \, \nonumber
\\
\label{eq:12}
\end{eqnarray}

with the Jacobian elliptic functions $sn(\varpi,m)$ and $
cn(\varpi,m)$ defined in Eq.~(\ref{eq:55}). Then, for any two
points such that $\varpi_2=\varpi_1 \mp 2K(m)$, one infers that:
$\dot{x}_{cl}(\varpi_2)=\,\dot{x}_{cl}(\varpi_1)$ and
$\ddot{x}_{cl}(\varpi_2)=\,\ddot{x}_{cl}(\varpi_1)$.  Thus,
periodic boundary conditions (PBC) apply to our problem.

In fact, being a product over an infinite number of eigenvalues
whose modulus is larger than one, $Det^R[\hat{O}]$ is
(exponentially) divergent in the $L\rightarrow \infty$ limit
whereas {\it ratios} of functional determinants are finite and
physically meaningful both in value and sign
\cite{gelfand,kleinert}. Therefore ${Det^R[\hat{O}]}$ has to be
normalized over the harmonic oscillator determinant
${Det[\hat{h}]}$ with $\hat{h}\equiv \, -\partial^2_{\tau} +
\omega^2$. For any two points separated (as above) by $2K(m)$, the
latter corresponding to the period $L$ along the $\tau-$axis, the
determinants ratio is formed in the case of PBC by
\cite{kleinert1}:

\begin{eqnarray}
{Det[\hat{h}]}&=&\,-4\sinh^2(\omega L/2)\, \nonumber
\\
{{Det^R[\hat{O}]} \over {<f_0 |f_0>}}&=&\,{{f_1(\varpi_2) -
f_1(\varpi_1)} \over {{f_0(\varpi_1) W(f_0, f_1)}}}\, \nonumber
\\
f_1&=&\,{{\partial {x}_{cl}(\tau)} \over {\partial m}}\, \,
\nonumber
\\
f_0&=&\,\dot{x}_{cl}(\tau)\, {}\, \label{eq:13}
\end{eqnarray}

where $f_0$, $f_1$ are independent solutions of the homogeneous
equation associated to Eq.~(\ref{eq:26}), $W(f_0, f_1)$ is their
Wronskian and $<f_0|f_0>$ is the squared norm which can be
computed by the first in Eq.~(\ref{eq:12}).

Working out the analytical calculation for the second in
Eq.~(\ref{eq:13}), I find:

\begin{eqnarray}
{{Det^R[\hat{O}]} \over {<f_0 |f_0>}}=\,{{(2 - m^2)^{7/2}}\over
{a^2 \omega^3 m^4}} \Biggl[ {{E(\pi/2,m)}\over {1 - m^2}} - K(m)
\Biggr] \label{eq:17}
\end{eqnarray}

where $E(\pi/2,m)$ is the complete elliptic integral of the second
kind. As the squared norm has dimension $[\omega a^2]$,
$Det^R[\hat{O}]$ correctly carries the dimension $[\omega^{-2}]$
consistently with the fact that $Det[\hat{O}]$ is dimensionless.

In the $E\rightarrow 0$ ($L \rightarrow \infty$) limit, $m^2
\rightarrow 1$ and $K(m) \sim \ln(4/\sqrt{1 - m^2})$. Moreover,
$<f_0 |f_0> \sim {4 \over 3}\omega a^2$.

Then, from Eqs.~(\ref{eq:13}),~(\ref{eq:17}), in the same limit, I
get

\begin{eqnarray} {{Det[\hat{h}]} \over
{Det^R[\hat{O}]}} \rightarrow \,- {{12\omega^2}} \label{eq:24}
\end{eqnarray}

\begin{figure}
\includegraphics[height=7cm,angle=-90]{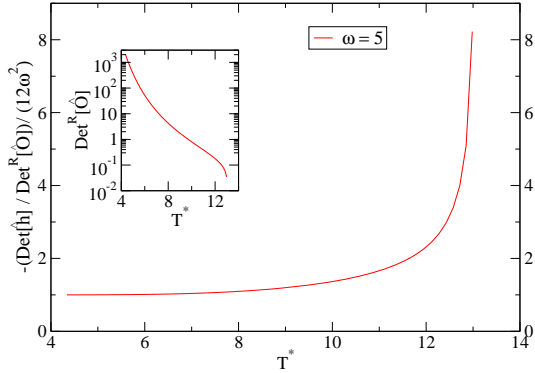}
\caption{\label{fig:3}(Color online) Normalized determinants ratio
versus $T^* (K)$. The inset plots the regularized fluctuation
determinant (units $meV^{-2}$) of Eq.~(\ref{eq:17}). }
\end{figure}

Thus recovering the well known result of the {\it infinite size}
bounce theory. In fact this occurs quite away from the zero limit
as clearly shown in Fig.~\ref{fig:3} where the determinants ratio
(normalized over $-12\omega^2$) is plotted against $T^*$ obtained
from Eq.~(\ref{eq:110}). As $\omega=\,5meV$, $T_{max}^*$ is set
here at $\sim 13K$ although it may be much larger in real systems
\cite{hanggi}. The result of Eq.~(\ref{eq:24}) is already achieved
at $T^* \sim 7K$, that is about the value at which the {\it finite
size} bounce fits the {\it infinite size} bounce (see
Fig.~\ref{fig:2}). This looks remarkable as it suggests that
quantum fluctuations and classical path start to experience the
{\it finite size} effects starting from the same temperature. In
the $T^*\rightarrow 0$ limit, computation of Eq.~(\ref{eq:17})
becomes time consuming in order to control the elliptic integrals
and reproduce the exponential divergence due to the $(1 -
m^2)^{-1}$ term.

The negative sign in Eq.~(\ref{eq:24}) is crucial and occurs for
any $T^*$. Then, the square root of the determinant ratio (which
appears in the partition function) is imaginary. As stated at the
beginning and made clear in Fig.~\ref{fig:2} the bounce has always
a maximum hence, the zero mode is not the ground state: there is a
negative eigenvalue $\varepsilon_{-1}$ \cite{n5} embedded in
$Det^R[\hat{O}]$ which causes a finite lifetime for the particle
placed at $x=\,x_1$. While such eigenvalue is easily found
($\varepsilon_{-1}=\,-3\omega^2$) in the {\it infinite size}
theory as the operator in the l.h.s. of Eq.~(\ref{eq:26}) becomes
of Rosen-Morse type \cite{landau}, here I am facing a non trivial
problem: that to determine $\varepsilon_{-1}$ at finite $E$ and
establish its effect in the temperature dependence of the particle
lifetime.

\section*{B. Lam\`{e} equation}

To pursue this goal, insert the last of Eqs.~(\ref{eq:25}) in the
stability equation ~(\ref{eq:26}) which easily transforms into:

\begin{eqnarray}
& &{{d^2} \over {d\varpi^2}} \eta_n(\tau) =\,\bigl[ l(l + 1)m^2
sn^2(\varpi,m) + \mathcal{A}_n \bigr] \eta_n(\tau) \, \nonumber
\\
& &\mathcal{A}_n=\,{1 \over {\chi_2^2}} - l(l + 1) -
{{\varepsilon_n}\over {(\chi_2 \omega)^2}}\,\nonumber \\ & &l(l +
1)=\,6 \label{eq:28}
\end{eqnarray}

This is the Lam\`{e} equation in the Jacobian form
\cite{whittaker} that, for given $l$ and $m$, admits periodic
solutions (which can be expanded in infinite series) for an
infinite sequence of characteristic $\mathcal{A}_n$ values. The
continuum of the fluctuation spectrum stems from this sequence.
However, for positive and integer $l$, (in the case at hand,
$l=\,2$), the first $2l + 1$ solutions of Eq.~(\ref{eq:28}) are
not infinite series but polynomials in the Jacobi elliptic
functions with real period $2K(m)$ or $4K(m)$ \cite{ward}. $2K(m)$
plays the role of the lattice constant being the period of the
potential in the Schr\"{o}dinger like stability equation. It is
among these five polynomial solutions corresponding to the
eigenvalues $\mathcal{A}_n$ that one has to pick up the unstable
fluctuation eigenstate.  Two solutions out of five have to be
discarded because they do not fulfill the periodicity conditions
required for the fluctuation components:
$\eta_n(\varpi)=\,\eta_n(\varpi \mp 2K(m))$. The three
(unnormalized) good solutions are \cite{i3}:

\begin{eqnarray}
& &\eta_{-1} \propto \,sn^2(\varpi,m) - {1 \over {1 + m^2 -
\sqrt{m^4 - m^2 +1}}} \, \nonumber
\\
& &\eta_0 \propto \,sn(\varpi,m)cn(\varpi,m)  \, \nonumber
\\
& &\eta_{1} \propto \,sn^2(\varpi,m) - {1 \over {1 + m^2 +
\sqrt{m^4 - m^2 +1}}} \, \nonumber
\\
\label{eq:32}
\end{eqnarray}

and, using Eqs.~(\ref{eq:28}), I get the corresponding energy
bands $\varepsilon_n$ as functions of the modulus $m$:

\begin{eqnarray}
& & \varepsilon_{-1}=\,-\omega^2\Biggl(1 + {2 \over {2 - m^2}}
\sqrt{m^4 - m^2 +1}\Biggr) \, \nonumber
\\
& &\varepsilon_0=\,0 \, \nonumber
\\
& &\varepsilon_{1}=\,-\omega^2\Biggl(1 - {2
\over {2 - m^2}} \sqrt{m^4 - m^2 +1}\Biggr) \label{eq:31}
\end{eqnarray}

\begin{figure}
\includegraphics[height=7cm,angle=-90]{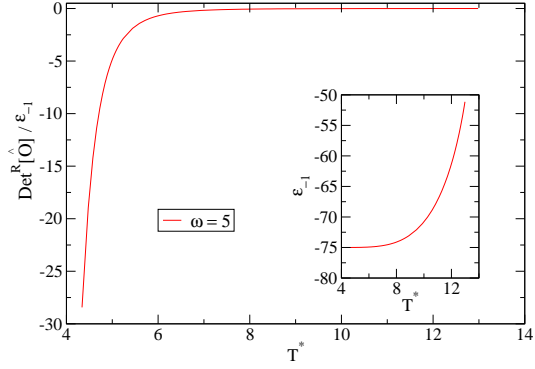}
\caption{\label{fig:4}(Color online) Ratio (in $meV^{-4}$) between
the regularized fluctuation determinant and the ground state
eigenvalue versus temperature. The inset plots separately the
$T^*$ dependence of the g.s. eigenvalue (in $meV^{2}$).}
\end{figure}

The zero mode is thus consistently recovered also for the {\it
finite size} theory with $\eta_0 \propto \dot{x}_{cl}$ \cite{n4}.
$\eta_1$ lies in the continuum with $\varepsilon_{1}\rightarrow
\omega^2$ in the {\it infinite size} limit. $\varepsilon_{-1}$ is
the object of our focus. Through Eqs.~(\ref{eq:5}),~(\ref{eq:8})
and ~(\ref{eq:110}), the $T^*$ dependence of $\varepsilon_{-1}$ is
computed as shown in the inset of Fig.~\ref{fig:4}: the value
$-3\omega^2$ of the {\it infinite size} theory is mantained up to
$T^*\sim 7K$ while $|\varepsilon_{-1}|$ softens significantly at
larger $T^*$. This strongly affects the overall behavior of
$Det^R[\hat{O}]$ as I point out by plotting the ratio
$Det^R[\hat{O}]/\varepsilon_{-1}$ on a linear scale.

\begin{figure}
\includegraphics[height=7cm,angle=-90]{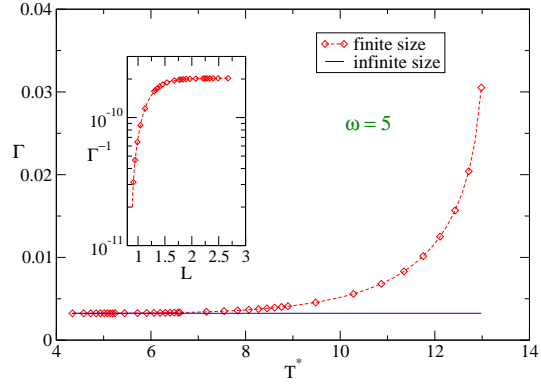}
\caption{\label{fig:5}(Color online) Tunneling rate (in $meV$)
computed from Eq.~(\ref{eq:54}) versus temperature. The rate of
the {\it infinite size} theory is plotted for comparison. The
inset shows the lifetime (in $sec$) against the {\it size} $L$.}
\end{figure}

\section*{IV. Decay Rate }

All the ingredients are now available to compute Eq.~(\ref{eq:25})
and derive the total partition function $Z_T$. Note that also the
solution of Eq.~(\ref{eq:1}), $x_{cl}=\,0$ contributes with a term
$Z_h$ to $Z_T$. Being $Z_h=\,\aleph |Det[\hat{h}]|^{-1/2}$, we now
fully understand on physical grounds the need to normalize
$Det^R[\hat{O}]$ over $Det[\hat{h}]$ as described above. In fact,
one has to account for all multiple excursions to and from the
abyss which is equivalent to sum over an infinite number of (non
interacting) bounce contributions as $Z_1$ in Eq.~(\ref{eq:25}).
The final result is: $Z_T=\,Z_h \exp(Z_1/Z_h)$. The tunneling rate
$\Gamma(L)$ is thus obtained from the imaginary part of the
exponent in $Z_T$ ($Z_1$ is purely imaginary) through the
Feynman-Kac formula \cite{schulman}. The general expression for a
finite size system is:

\begin{eqnarray}
& & \Gamma(L)=\,\hbar  \sqrt{{{A[x_{cl}] + E\cdot L}\over
{2\pi\hbar}}} \sqrt{{{Det[\hat{h}]} \over {Det^{RR}
[\hat{O}]\bigl|\varepsilon_{-1}\bigl| }}} \exp\biggl[- {1 \over
{\hbar}} A[x_{cl}] \biggr]  \, \nonumber
\\
\label{eq:54}
\end{eqnarray}

with $Det^{RR} [\hat{O}]\equiv \,Det^R[\hat{O}]/\varepsilon_{-1}$
being negative as shown in Fig.~\ref{fig:4}. Eq.~(\ref{eq:54}) is
plotted in Fig.~\ref{fig:5} against $T^*$ together with the
constant tunneling rate of the {\it infinite size} theory.
$\Gamma(T^*)$ grows fast as the length of the bounce shrinks and
the quantum fluctuations spectrum softens. Close to $T_{max}^*$,
the tunneling rate has increased by an order of magnitude with
respect to the {\it infinite size} result but the decay width
remains smaller than the fundamental oscillator energy being
$\Gamma(T^*=\,12.98K)/\omega \sim 6 \cdot 10^{-3}$. The inset in
Fig.~\ref{fig:5} plots the lifetime of the false vacuum versus the
system size.

\section*{V. Conclusion}

This work presents a study of the {\it finite size} effects on a
metastable quartic nonlinear potential and, as such, is
complementary to a previous investigation \cite{i1}. The
semiclassical path integral method, implemented by the theory of
the functional determinants, builds the framework to analyse the
tunneling in the metastable system. The description relies on the
properties of the elliptic functions which permit to monitor the
evolution of the classical bounce versus the system size. After
defining the temperature range within which the tunneling occurs,
I have computed the temperature effects on the quantum fluctuation
spectrum. In particular, the negative eigenvalue which causes
metastability has been studied in great detail solving a Lam\`{e}
type equation. These results, new in the literature, permit to
compute the physical properties of the false ground state.
Specifically I have shown that its lifetime has a remarkable
size/temperature dependence inside the quantum regime. While the
latter can persist up to $T_{max}^*$ of order $100K$ in real
systems, the presented quantitative method may also be of
practical interest provided the potential parameters are adapted
to specific cases.

\appendix*

\section{}

The Jacobian elliptic functions used throughout the paper are
related to the amplitude $\zeta$ in Eq.~(\ref{eq:8}) by the
definitions

\begin{eqnarray}
& &  sn(\varpi,m)=\,\sin\zeta \, \nonumber
\\
& &  cn(\varpi,m)=\,\cos\zeta \, \nonumber
\\
& &  dn(\varpi,m)=\,\sqrt{1 - m^2\sin^2\zeta}
\label{eq:55}
\end{eqnarray}

In the computation of the bounce and its time derivative, the
following representations as trigonometric series have been used
\cite{grad}

\begin{eqnarray}
& &sn(\varpi,m)=\,{{2\pi} \over {mK(m)}} \sum_{n=1}^{\infty}{{q^{n
- 1/2 }} \over {1 - q^{2n - 1}}} \sin \Bigl({{(2n - 1) \pi \varpi}
\over {2K(m)}} \Bigr)\,\nonumber
\\
& &cn(\varpi,m)=\,{{2\pi} \over {mK(m)}} \sum_{n=1}^{\infty}{{q^{n
- 1/2 }} \over {1 + q^{2n - 1}}} \cos \Bigl({{(2n - 1) \pi \varpi}
\over {2K(m)}} \Bigr)\,\nonumber
\\
& &dn(\varpi,m)=\,{{\pi} \over {2K(m)}} + {{2\pi} \over {K(m)}}
\sum_{n=1}^{\infty}{{q^{n}} \over {1 + q^{2n}}} \cos \Bigl({{n \pi
\varpi} \over {K(m)}} \Bigr)\,\nonumber
\\
& &q=\,\exp\bigl(-\pi K'(m)/K(m)\bigr)\, \nonumber
\\
& &K'(m)=\,F(\pi/2,\bar{m})\, \nonumber
\\
& &\bar{m}^2=\,{1 - m^2}
 \label{eq:56}
\end{eqnarray}

with $\varpi$ defined in Eq.~(\ref{eq:9}). Numerical convergence
is achieved by taking a cutoff $n_{max}\sim \,40$ in the Fourier
series.


\begin{references}
\bibitem{n1}
Note that the classical problem admits a solution also for
positive $E$ up to the sphaleron energy $V(x=\,a)$. However, in
this case there would be only one turning point located at $x >
\sqrt{2}a$. Thus the particle would explore portions of the abyss
which become broader by increasing $E$.
\bibitem{i2}
Sometimes the term "instanton" is misleadingly used to denote also
the "bounce" solution. The different boundary conditions fulfilled
by the two objects underlie profound differences in their
dynamics. The former interpolates between the potential minima
going across the barrier (the valley in the reversed potential)
nearly instantaneously and with finite velocity, hence it spends a
very short time around its center of motion $\tau_0$. The latter
starts from the false vacuum, travels through the barrier and
reaches the edge of the abyss at $\tau_0$. There it stops, inverts
the motion and goes back to the initial position. Then, the
Goldstone modes related to the instanton and bounce trajectories
have zero and one node respectively.
\bibitem{n3}
The dimensionalities are: $[\varepsilon_n] \equiv [sec^{-2}]$,
$[\eta_n(\tau)] \equiv [sec^{-1/2}]$, $[\varsigma_n] \equiv [\AA
\, sec^{1/2}]$.
\bibitem{n4}
The path velocity is proportional to the zero mode fluctuation
$\dot{x}_{cl}(\tau)=\,K\eta_0(\tau)$ and $K^{2}$ is the squared
norm $<f_0|f_0>$ used in Eq.~(\ref{eq:13}).
\bibitem{n4a}
Precisely, the replacement is: $(\varepsilon_0)^{-1/2} \rightarrow
\sqrt{{{A[x_{cl}]+ E\cdot L}\over {2\pi\hbar}}}L$ which tends to
the expression in the text for the $T\rightarrow 0$ limit.
\bibitem{n5}
In Eq.~(\ref{eq:34}) and following the sums and products run over
$n=\,0...\infty$. To be precise they all contain also the negative
eigenvalue formally defined by $\varepsilon_{(n=\,-1)}$.
\bibitem{i3}
Note that the way to order eigenvalues and eigenstates in terms of
$n$ is purely conventional. Either $n \in [-l,l]$ or $n \in [0,2l
+ 1]$ can be used. The former is assumed consistently with having
labelled the unstable eigenvalue by $n=\,-1$. See \cite{n5}.
\end{references}
\end{document}